\documentclass[10pt]{article}

\usepackage{amsmath}
\usepackage{amssymb}
\usepackage{graphicx}

\usepackage{cite}

\usepackage{color} 

\date{}

\newcommand{\beq}{\begin{equation}}
\newcommand{\eeq}{\end{equation}}
\newcommand{\bbar}{\begin{eqnarray}}
\newcommand{\eear}{\end{eqnarray}}

\begin{document}

%\linenumbers

% Title must be 150 characters or less
\begin{flushleft}
{\LARGE
\textbf{Symmetric vs asymmetric stem cell divisions: an adaptation against cancer?}
}
% Insert Author names, affiliations and corresponding author email.
\bigskip 
\bigskip 
\bigskip 
\\
Leili Shahriyari$^1$,
Natalia L. Komarova$^{1,\ast}$
\bigskip 
\bigskip 
\bigskip 
\\
\bf{1} Department of Mathematics, University of California Irvine, Irvine, CA, USA
\\
\bigskip 
\bigskip
 \bigskip 

$\ast$ E-mail: komarova@uci.edu, tel. +1-949-824-1268
\end{flushleft}

\bigskip

%Running title: Symmetric vs asymmetric stem cell divisions

\newpage 
% Please keep the abstract between 250 and 300 words
\section*{Abstract} 

Traditionally, it has been held that a central characteristic of stem
cells is their ability to divide asymmetrically. Recent advances in
inducible genetic labeling provided ample evidence that symmetric stem
cell divisions play an important role in adult mammalian
homeostasis. It is well understood that the two types of cell divisions
differ in terms of the stem cells' flexibility to expand when
needed. On the contrary, the implications of symmetric and asymmetric
divisions for mutation accumulation are still poorly understood.  In
this paper we study a stochastic model of a renewing tissue, and
address the optimization problem of tissue architecture in the context
of mutant production. Specifically, we study the process of tumor
suppressor gene inactivation which usually takes place as a sequence
of two consecutive "hits", and which is one of the most common
patterns in carcinogenesis. We compare and contrast symmetric and
asymmetric (and mixed) stem cell divisions, and focus on the rate at
which double-hit mutants are generated. It turns out that
symmetrically-dividing cells generate such mutants at a rate which is
significantly lower than that of asymmetrically-dividing cells. This
result holds whether single-hit (intermediate) mutants are
disadvantageous, neutral, or advantageous. It is also independent on
whether the carcinogenic double-hit mutants are produced only among
the stem cells or also among more differentiated cells. We argue that
symmetric stem cell divisions in mammals could be an adaptation which
helps delay the onset of cancers. We further investigate the question
of the optimal fraction of stem cells in the tissue, and quantify the
contribution of non-stem cells in mutant production. Our work provides
a hypothesis to explain the observation that in mammalian cells, symmetric
patterns of stem cell division seem to be very common. 

%The force of selection that comes from the cancer-delaying effect of such an architecture can be thought to have helped shape the observed structures.

% Please keep the Author Summary between 150 and 200 words
% Use first person. PLoS ONE authors please skip this step. 
% Author Summary not valid for PLoS ONE submissions.   
\section*{Author Summary}

Most cells divide symmetrically, by creating two copies of
themselves. Stem cells have been considered an exception from this
rule, because of their ability to divide in an asymmetric fashion, to
create one stem and one differentiated cell. This asymmetric division
mechanism became associated with the very concept of "stemness". In
contrast to the traditional view, recent evidence suggests that
asymmetric divisions, although common in some tissues, are less common
in others. There, stem cell divide in a symmetric fashion, by either
creating two copies of themselves, or two differentiated cells. What
are the advantages and disadvantages of different types of
self-renewal? One answer can come from examining evolutionary
consequences of various stem cell division patterns.  By using a
mathematical model of cellular turnover in tissues, we studied the
ability of cells to produce double-hit mutants (such mutants are often
responsible for cancer initiation). It turns out that
symmetrically-dividing cells generate such mutants at a rate which can
be orders of magnitude lower than that of asymmetrically-dividing
cells. This argument provides an evolutionary hypothesis of why stem
cells in mammals often divide symmetrically.

\section*{Introduction}

The ability of stem cells to divide asymmetrically to produce one stem
and one non-stem daughter cell is often considered to be one of the
defining characteristics of stemness. On the other hand, there is ample evidence suggesting that
adult stem cell can and do divide symmetrically \cite{morrison2006asymmetric, shen2004endothelial}. 

\begin{figure}[!ht]
\begin{center}
\includegraphics[width=4in]{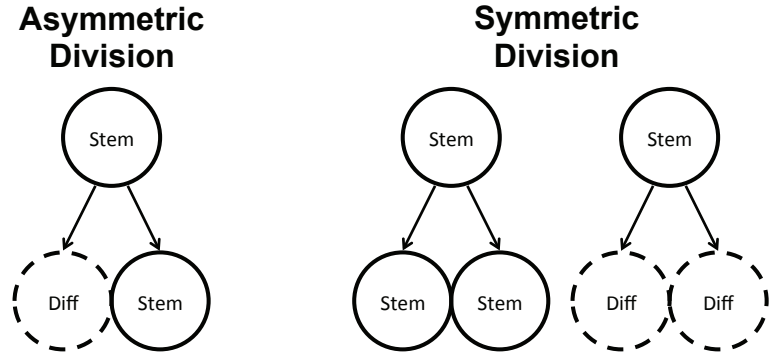}
\end{center}
\caption{
{\bf Symmetric and asymmetric stem cell divisions.}   In the asymmetric division model, a stem cell produces one differentiated cell and one stem cell. In the symmetric division model, a stem cell produces two differentiated cells or two stem cells.
}
\label{fig:SymAsym}
\end{figure}

Two basic models of stem cell divisions are discussed in the
literature, see figure \ref{fig:SymAsym}. The asymmetric model
suggests that the homeostatic control of the stem cell pool is
maintained at the level of single cells, whereby each stem cell
produces a copy of itself plus one differentiated cell
\cite{knoblich2008mechanisms, fuchs2004socializing,
  zhong2008neurogenesis, ho2005kinetics}. From the engineering
prospective, this model has the advantage of keeping the stem cell
population level steady. An obvious disadvantage is its inability to
replenish the stem cell pool in case of injury. This problem is
naturally solved by the symmetric model, which maintains homeostatic
control at the population level, rather than at the individual cell
level. There, stem cells are capable of two types of symmetric
divisions: a proliferation division resulting in the creation of two
stem cells, and a differentiation division resulting in the creation
of two differentiated cells \cite{zhang2009distinct,
  loeffler2002tissue,
  marshman2002intestinal,clayton2007single}. Differentiation/proliferation
decisions are though to be under control of numerous signals emanating
from the surrounding tissue and the stem cells themselves
\cite{liu2000loss,simmons2003cyclic,alvarez2004long,saha2006inhibition,lien2006alpha,adams2007niche,dehay2007cell,
  orford2008deconstructing, nusse2008wnt, spiegel2008stem,
  saha2008tgf,sen2008mechanical, guilak2009control, lavado2010prox1,
  de2010regulation,li2010coexistence,
  salomoni2010cell,hsieh2012orchestrating, ordonez2012lrig1}.

Uncovering division patterns of stem cells has been subject of intense
research in the last fifteen years. Some of the first quantification
of the division strategies in vitro comes from the work of
\cite{yatabe2001investigating} who tracked methylation patterns in the
dividing cells of the colon crypts. The analysis of the complex
methylation patterns revealed that crypts contain multiple stem cells
that go through ``bottlenecks'' during the life of the organism, which
suggests that symmetric divisions are part of the picture. Another
piece of evidence comes from experiments with chimeric mice to
determine the dynamics of polyclonality of crypts. Initially
polyclonal crypts eventually become monoclonal, which suggests that
symmetric divisions must occur \cite{spradling2001stem, nicolas2007stem}. By means of
radiotherapy-induced mutations, the study of \cite{campbell1996post}
suggests that a significant fraction of the somatic mutations in human
colon stem cells are lost within one year.

An important advance in quantification of symmetric vs antisymmetric divisions
became possible with the invention of inducible genetic labeling
\cite{klein2011universal}. This technique provides access to
lineage-tracing measurements, from which the fate of 
labeled cells and their clones can be tracked over time. By
means of the quantitative analysis of long-term lineage-tracing data
\cite{klein2010mouse, clayton2007single}, it has been shown that the
rate of stem cell replacement is comparable to the cell division rate,
implying that  symmetric cell divisions contribute significantly 
to stem cell homeostasis \cite{lopez2010intestinal,
  snippert2010intestinal}. The paper by \cite{simons2011strategies}
provides a review of the recent evidence of symmetric divisions in
mammalian intestinal stem cells, spermatogenesis \cite{klein2010mouse}
and epithelial tissues such as hair follicles \cite{doupe2010ordered}.

These new findings reveal that contrary to the previous thinking, adult tissue stem cells are often lost (e.g. by differentiation) 
and replaced in a stochastic manner. This notion challenges the traditional
concept of the stem cell as an immortal, slow-cycling, asymmetrically
dividing cell \cite{klein2011universal}.

In paper \cite{simons2011strategies}, an important question is
raised: Why should mechanisms of tissue maintenance so often lean
toward symmetric self-renewal? One answer comes from
recognizing the ability of symmetrically-dividing stem cells to
respond to injury thus ensuring a robust mechanism of tissue
homeostasis. It however could be argued that the symmetric divisions are  "switched on" in response to a sudden stem cell loss, and the asymmetric division strategy is employed in the course of normal homeostasis.

In the present paper, we explore an alternative hypothesis, which
gives an additional ``reason'' for the tissue architecture favoring
symmetric divisions. As a starting point, we note that in both
symmetric and asymmetric division types, a dysregulation may lead to
the loss of homeostatic control and an unchecked growth of cells. A
disruption in the control of proliferation/differentiation decisions
can tip the balance and lead to abnormal stem cell expansion
\cite{reya2005wnt}. It has also been shown that disruption of
asymmetric divisions can be responsible for cancerous growth of
undifferentiated cells \cite{clarke2003regulation,
  caussinus2005induction, caussinus2007asymmetric, Aparicio2009,
  gonzalez2013drosophila}.

Here, we examine the symmetric and asymmetric divisions in the context
of producing mutations. Many cancerous transformations start off by an
inactivation of a tumor-suppressor gene
\cite{vogelstein2002genetic}. This is the famous two-hit process
discovered by Knudson \cite{knudson1971mutation, knudson2001two} and
studied by many laboratories as well as theoretically.  We ask the following question: from the point of view of two-hit
mutant generation, what type of stem cell divisions is advantageous
for the organism? What frequency of symmetric vs asymmetric divisions
can maximally delay the stochastic generation of a dangerous mutant?
To this end, we consider a continuous range of strategies with mixed type
divisions and explore how the frequency of symmetric vs asymmetric
divisions affects the generation of mutations.

In this paper, we use both numerical simulations and analytical
methods to study symmetric and asymmetric stem cell divisions in the
context of mutation production. Other theoreticians have explored stem
cell dynamics by means of deterministic stem cell modeling and
stochastic numerical simulations \cite{pmid7479951, pmid17049944,
  pmid17360468, pmid18451157, pmid12101397, pmid11517339,
  ganguly2006mathematical, ganguly2007mathematical,
  boman2007symmetric, ashkenazi2008pathways, michor2008mathematical,
  tomasetti2010role, enderling2011cancer, enderling2007mathematical,
  enderling2009paradoxical, enderling2009importance,
  enderling2009migration}. A great review of many modeling approaches
is provided in \cite{piotrowska2008mathematical}. \cite{frank2003cell}
studied the dynamics of mutation spread in development and showed that
susceptibility to late-life cancers may be influenced by somatic
mutations that occur during early
development. \cite{michor2003stochastic} considered a model of stem
cell dynamics, and calculated the rates of stochastic elimination (or
washing out) of mutants. In this model, stem cells can proliferate
symmetrically and differentiation is decoupled from proliferation. 
\cite{dingli2007symmetric} considered the question of mutation
generation by stem cells and found that mutations that increase the
probability of asymmetric replication can lead to rapid expansion of
mutant stem cells in the absence of a selective fitness advantage.

In the present paper, we concentrate on the optimization problem of
tissue architecture in the context of delaying double-hit mutant
production, and focus specifically on symmetric and asymmetric stem cell divisions. We consider a stochastic model of double-hit mutant
generation, and ask several questions related to evolutionary dynamics
of mutations. What type of divisions is optimal? What cell types
contribute the most to double-hit mutant generation? What is the
optimal fraction of stem cells that delays carcinogenesis?

\section*{Results}

\subsection*{Set-up}

We consider a two-compartment, agent-based model of stem cells and
transit-amplifying (TA) cells. The stem cells are capable of both
symmetric and asymmetric divisions (see figure \ref{fig:SymAsym}). The
relative proportion of symmetric divisions can vary and is denoted by
the symbol $\sigma$ (see Table \ref{tab:1}), where $\sigma=1$ means
that all divisions are symmetrical, and $\sigma=0$ means that stem
cells only divide asymmetrically. The symmetric divisions can be of
two types, proliferation and differentiation. The type of symmetric
division is defined by a regulatory mechanism which assures an
approximately constant level of stem cells (see Methods). The total population
(which includes both stem cells, $S$, and TA cells, $D$) is
denoted by $N=S+D$. An important parameter is $\lambda=N/D$, which
defines the proportion of stem cells with respect to TA
cells: $S/D=\lambda-1$.

\begin{table}[!ht]
\caption{
\bf{Model parameters}}
\begin{tabular}{|c|c|}
\hline
Notation & Description\\
\hline
$S,D$ & Number of stem and non-stem cells \\
$N=S+D\gg 1$ & Total population size\\
$\lambda=N/D>1$ & Inverse relative number of non-stem cells \\
$0\le \sigma\le 1$ & Proportion of symmetric stem cell divisions\\
$r\lesssim 1$, $r\gtrsim 1$  & Fitness of one-hit mutants \\
$u_1,u_2\ll 1$ & Mutation rates leading to the acquisition of first and second hits\\
%$j_*$, $j$ & The number of stem and non-stem mutant cells\\
$R_{0\to 2}$ & Tunneling rate\\
\hline
\end{tabular}
\begin{flushleft}Notations used in the text and their brief description. 
\end{flushleft}
\label{tab:1}
\end{table}

We assume that the non-stem cells die, and that all cell types have a
chance to divide. Each time a division happens, there is a
probability, $u_1$, that one of the daughter cells is a one-hit mutant
with fitness $r$ (while the fitness of all wild-type cells is given by
$1$). The fitness parameter defines the relative probability of the
given cell-type to be chosen for division. In this paper we consider a
range of fitness values, $r$, such that the one-hit mutants can be
disadvantageous compared to wild-type cells,  neutral, or
even slightly advantageous. When a one-hit mutant divides, it has
the probability $u_2$ to give rise to a two-hit mutant. Two-hit
mutants are transformed cells which have a potential to give rise to a
cancerous tissue transformation.

The generation of two-hit mutants is normally considered to be a
rate-limiting step in cancer initiation. Once such a mutant is
produced, it may break down homeostatic control and result in a wave
of clonal expansion, followed by further transformations. It is this
first step, the creation of a double-hit mutant, that we focus on in
this paper. We investigate how the timing of such a mutant production
depends on the tissue architecture, and specifically, on the symmetry
of stem cell divisions. 

In order to gain analytical insights, a slightly simplified stochastic
process was considered (see the Methods Section) which gave
predictions that are in excellent agreement with the computational
model. 

\subsection*{Tunneling rates}

\begin{figure}[!ht]
\begin{center}
\includegraphics[width=6in]{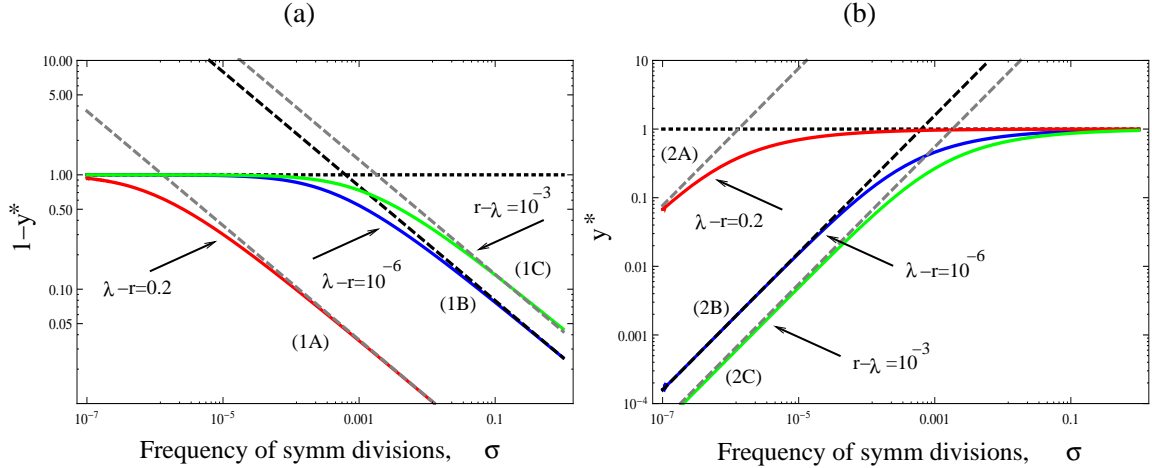}
\end{center}
\caption{
{\bf The six different approximation regimes (Table \ref{tab:2}) for solutions of system (\ref{Y1}-\ref{Y2}). }  Plotted is the quantity (a) $1-y_*$ and (b) $y_*$ as a function of the frequency of symmetric divisions, $\sigma$, for three different values of $\lambda$ (solid lines), together with the approximations given by the formulas in Table \ref{tab:2}. Approximations $(1A)$, $(1B)$, and $(1C)$ are best demonstrated in panel (a), where the quantity $1-y_*$ is plotted. Approximations $(2A)$, $(2B)$, and $(2C)$ are best demonstrated in panel (b), where the quantity $y_*$ is plotted. The other parameters are $u_1=u_2=10^{-7}$, $r=1.1$.
}
\label{fig:app}
\end{figure}

While the detailed temporal dynamics of double-mutant production is
given in the Methods Section, here we present the results for the
so-called ``tunneling rates'' - the rates at which the stem cell
system of a given size produces double-hit mutants (assuming that
one-hit mutants drift at relatively low levels). Denoting the
tunneling rate as $R_{0\to 2}$ (where the subscript suggests that the
system transfers from all wild-type, ``zero-hit'', state to a system
containing two-hit mutants), we have
\beq
\label{rate}
R_{0\to 2}=R^{stem}_{0\to 2}+R^{TA}_{0\to 2}=\frac{Nu_1}{2}\left[\left(1-\frac{1}{\lambda}\right)(1-y_*)+\left(1+\frac{1}{\lambda}\right)(1-y)\right],
\eeq
where quantities $y$ and $y_*$ satisfy the system
\bbar
\label{Y1}
0&=&\left[\frac{r\sigma}{2}(y_*^2+y^2)+r(1-\sigma)y_*y\right](1-u_2)-ry_*,\\
\label{Y2}
0&=&r(1-u_2)y^2+\lambda(1-y) -ry.
\eear
The time to produce double-hit mutants is distributed exponentially with the mean 
$$T_{0\to 2}=\frac{1}{R_{0\to 2}}.$$
Formula (\ref{rate}) describes the generation of double-hit mutants in
the stem cells (the first term on the right) and in TA cells
(the second term of the right). Several limiting cases are presented
in Table \ref{tab:2} and illustrated in figure \ref{fig:app}. 

\begin{table}[!ht]
\caption{
\bf{Important limiting cases for the tunneling rate  (formula (\ref{rate})).}}
\begin{tabular}{|c|c|c|c|c|}
\hline
Regime & Description & Conditions & $1-y_*$ & $1-y$ \\
\hline
$(1A)$ & $r<\lambda$, symm+asymm  & $\sigma\gg u_2$, $|\lambda-r|\gg \sqrt{u_2}$, $r<\lambda$ & $\sqrt{\frac{2u_2}{\sigma(1-r/\lambda)}}$ & $\frac{ru_2}{\lambda-r}$\\
$(1B)$ & $r\approx \lambda$, symm+asymm  &$\sigma\gg \sqrt{u_2}$, $|\lambda-r|\ll \sqrt{u_2}$ & $\sqrt{\frac{2\sqrt{u_2}}{\sigma}}$ & $\sqrt{u_2}$ \\
$(1C)$ & $r>\lambda$, symm+asymm  &$\sigma\gg |\lambda-r|$, $|\lambda-r|\gg \sqrt{u_2}$, $r>\lambda$ & $\sqrt{\frac{2}{\sigma}\left(\frac{r}{\lambda}-1\right)}$ & $1-\frac{\lambda}{r}$ \\
$(2A)$ & $r<\lambda$, asymm  & $\sigma\ll u_2$, $|\lambda-r|\gg \sqrt{u_2}$, $r<\lambda$ & $1-\frac{\sigma(1-r/\lambda)}{2u_2}$ & $\frac{ru_2}{\lambda-r}$\\
$(2B)$ &$r\approx \lambda$, asymm  & $\sigma\ll \sqrt{u_2}$, $|\lambda-r|\ll \sqrt{u_2}$ & $1-\frac{\sigma}{2\sqrt{u_2}}$ & $\sqrt{u_2}$ \\
$(2C)$ & $r>\lambda$, asymm  &$\sigma \ll |\lambda-r|$, $|\lambda-r|\gg \sqrt{u_2}$, $r>\lambda$ & $1-\frac{\sigma}{2(r/\lambda-1)}$ & $1-\frac{\lambda}{r}$ \\
\hline
\end{tabular}
\begin{flushleft} The notations for the six different regimes refer to figure \ref{fig:app}. 
\end{flushleft}
\label{tab:2}
\end{table}

%Comment on $\lambda=r$ condition??

Predictions of formula (\ref{rate}), as well as the more precise
equation (\ref{P2new}), have been compared with stochastic numerical
simulations, and found to be in excellent agreement with them, see below.

\subsection*{Double-hit mutants are produced slower under symmetric compared to asymmetric divisions.} 

\begin{figure}[!ht]
\begin{center}
\includegraphics[width=4in]{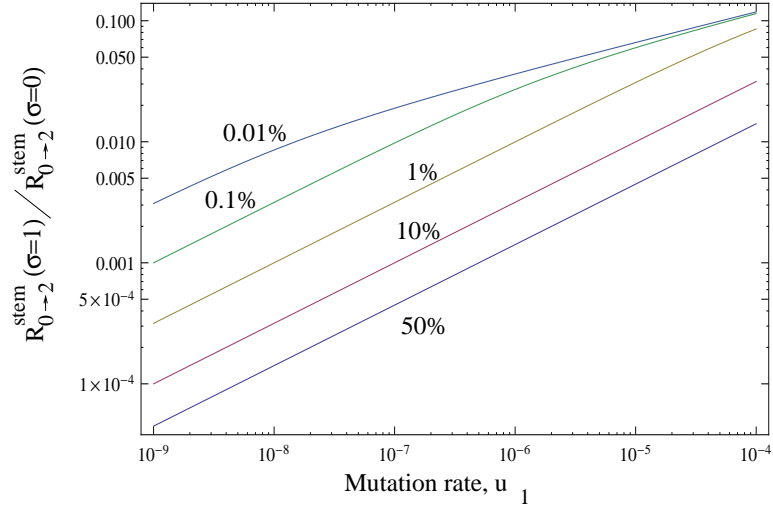}
\end{center}
\caption{
{\bf The reduction in the rate of double mutant production in stem cells with symmetric divisions compared to stem cells with asymmetric divisions only.} Plotted is the quantity in formula (\ref{quo}) as a function of the mutation rate, $u_1$. The percentage of the stem cells in the whole population ($S/N$) is marked next to the lines. The other parameters are $u_2=u_1/2$, $r=1$.
}
\label{fig:quo}
\end{figure}

An important question is how the fraction of symmetric divisions
($\sigma$) affects the rate of double-mutant production. We can see
that the production of double-mutants by non-stem cells does not
depend on $\sigma$, the frequency of symmetric divisions. On the other
hand, the production by stem cells is crucially affected by this
parameter.  Our formulas show clearly that the rate of tunneling grows
as $\sigma$ decreases, and it is the highest when $\sigma=0$, the case
of purely asymmetric divisions. This means that in order to minimize
the rate of double-hit mutant formation, one needs to maximize the
share of symmetric divisions. In figure \ref{fig:quo} we plot the quantity 
\beq
\label{quo}
\frac{\mbox{Rate of double-hit mutant production by stem cells, under symmetric divisions}}{\mbox{Rate of double-hit mutant production by stem cells, under asymmetric divisions}}=\frac{R^{stem}_{0\to 2}(\sigma=1)}{R^{stem}_{0\to 2}(\sigma=0)},
\eeq
for different percentages of stem cells. We can see that for realistic
ranges of the mutation rates, the difference is at least $10$-fold,
and can be as high as $10^4$-fold, with the symmetrically dividing
stem cells producing double-hit mutants slower than asymmetrically
dividing cells.

\begin{figure}[!ht]
\begin{center}
\includegraphics[width=6.9in]{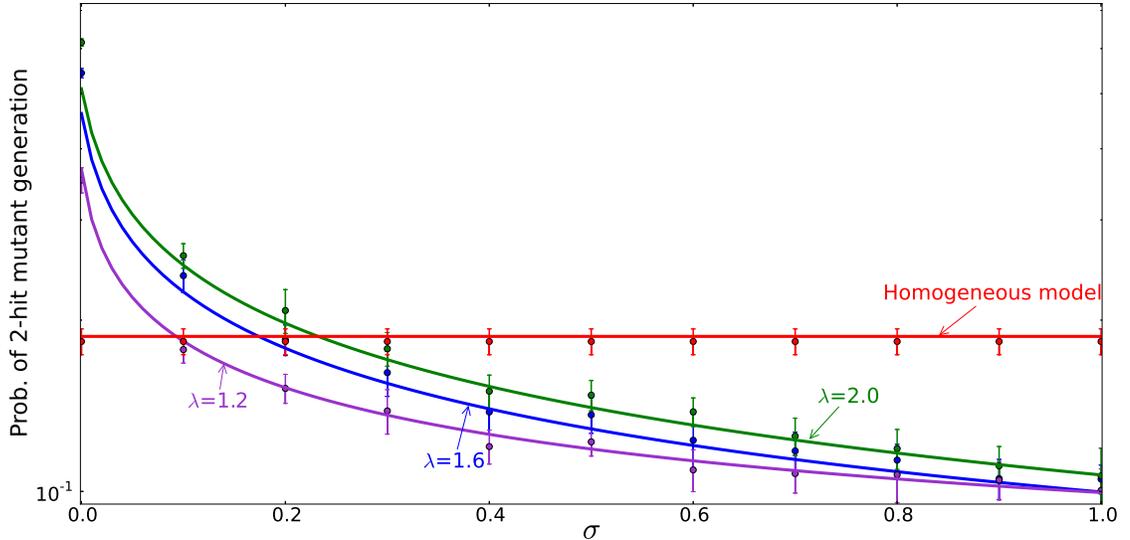}
\end{center}
\caption{
{\bf The probability of double-hit mutant generation as a function of $\sigma$, the probability of symmetric stem cell divisions.} 
The results of numerical simulations are presented as points connected with dotted lines (standard deviations are included).  Analytical results are given by solid lines (formula (\ref{P2new})). The horizontal line represents the calculations for the homogeneous model. 
We ran $10$ batches  of $1000$ runs. 
The parameters are $r=1.0$, $u_1=0.00001$, $u_2=0.002$, $N=500$. }  
\label{fig:DifferentSigma}
\end{figure}

Figure  \ref{fig:DifferentSigma} compares the analytical findings for the double-hit mutant production dynamics with the numerical simulations. We ran the stochastic numerical model (see Methods) for a fixed number of time-steps, and recorded whether or not a double-hit mutant has been generated. Repeated implementation of this procedure produced a numerical approximation of the probability of double-hit mutant generation, which is plotted (together with the standard deviations) as a function of $\sigma$, the probability of symmetric divisions, for three different values of $\lambda$, which measures the fraction of stem cells.  Clearly, the probability of mutant generation in the course of a given time-interval is a decaying function of $\sigma$. 

Another result that follows from our computations is the comparison of
the double-mutant production in a hierarchical (stem cells plus TA
cells) model compared with the conventional, homogeneous model that
has been extensively studied \cite{nowak2002role,
  komarova2003mutation, iwasa2004stochastic, weissman2009rate}. It
turns out the hierarchical model with purely asymmetric divisions
always produces mutants faster than the homogeneous model. For the
hierarchical model with purely symmetric divisions the result depends
on the fitness of one-hit mutants. For disadvantageous one-hit mutants
whose fitness satisfies $r<1$, $|1-r|\gg \sqrt{u_2}$, the hierarchical
model with purely symmetric divisions produces double-mutants faster,
and for neutral and advantageous mutants, it produces double-hit
mutants slower than the homogeneous model. In figure
\ref{fig:DifferentSigma} we can see that for $r=1$ (neutral one-hit
mutants), hierarchical models with a sufficiently large values of
$\sigma$ are characterized by slower double-hit mutant generation
compared to the homogeneous model (the horizontal line).

\begin{figure}[!ht]
\begin{center}
\includegraphics[width=6.9in]{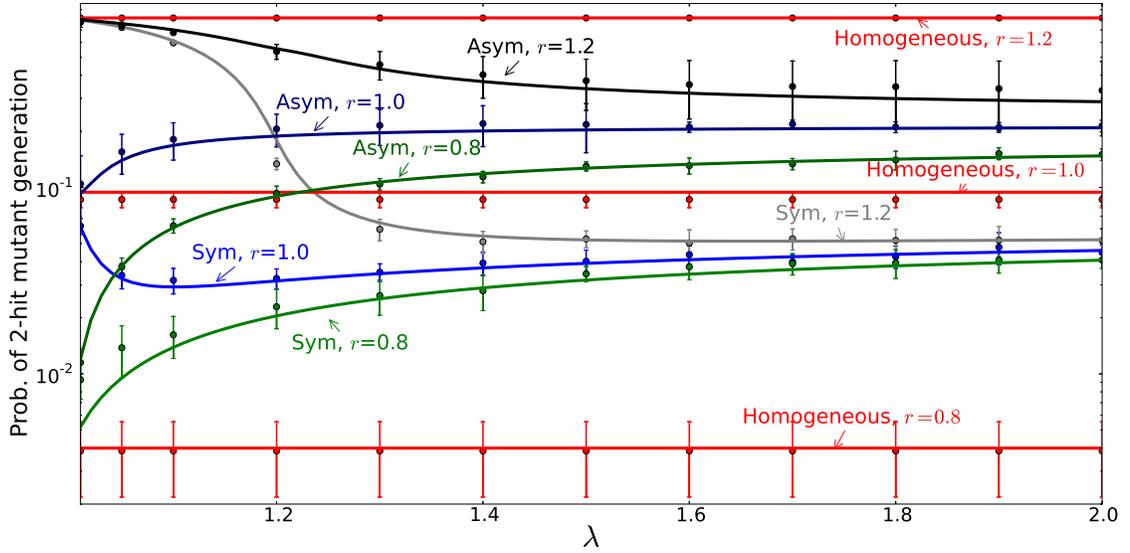}
\end{center}
\caption{
{\bf The probability of double-mutant generation as a function of $\lambda$, the ratio of TA cells to the total number of cells.} As in figure \ref{fig:DifferentSigma}, the results of numerical simulations are presented as points connected with dotted lines (standard deviations are included), and the analytical results are given by solid lines (formula (\ref{P2new})). The horizontal lines represent the calculations for the homogeneous model. 
We ran $10$ batches  of $1000$ runs. 
Plotted is the probability of double-mutant generation  as a function 
of $\lambda$, for purely symmetric ($\sigma=1$) and purely asymmetric ($\sigma=0$) models, for three different values of $r$.  The parameters are $u_1=10^{-5},  u_2=10^{-4},  N=1000$.}  
\label{fig:DifferentR}
\end{figure}

Figure \ref{fig:DifferentR} shows additional results of simulations
(together with our analytical calculations), where for three different
values of $r$ (one-hit mutant fitness) the probability of double-hit
mutant generation is plotted as a function of $\lambda$. The values
$\lambda\to 1$ corresponds to a vanishingly low fraction of stem cells
in the system, while $\lambda=2$ corresponds to $50\%$ of all cells
being stem cells. We show purely symmetric ($\sigma=1$) and purely
asymmetric ($\sigma=0$) cases. For fixed mutation rates and
populations sizes, the homogeneous model is characterized by only one
parameter, $r$, which is the fitness of one-hit mutants. The
probability of double-hit mutant generation strongly depends on whether
these intermediate mutants are disadvantageous ($r<1$), neutral
($r\approx 1$), or advantageous ($r>1$). In contrast to the homogeneous
model, the hierarchical model contains two additional parameters,
$\lambda$ (the ratio of TA cells and the total population) and
$\sigma$ (the probability of symmetric divisions). We can see that
these two parameters affect the probability of double-hit mutant
generation at least as strongly as the fitness $r$ does. The influence
of $\sigma$ is clear: the more the fraction of symmetric divisions, the
slower double-hit mutants are produced. Next, we examine the role of
the stem cell to TA cell ratio.

%Since the tunneling rate of both symmetric and asymmetric division will converge to the homogenous model as $\lambda \to 1$, this implies for $r>1$, the worst $\lambda$'s are those which are close to the one. However, for $r<1$ the smallest $\lambda$'s are the best ones. 

\subsection*{The optimal fraction of stem cells.} 

Let us consider an optimization problem for the tissue design,
with the goal to delay the production of double-hit mutants. What is
the optimal fraction of stem cells that the population should
maintain? Analysis of the tunneling rates for a hierarchical model
with purely symmetric divisions suggests that the optimal fraction of
stem cells depends on the fitness of the one-hit mutants. If the
one-hit mutants are disadvantageous ($r<1$, $|1-r|\ll \sqrt{u_2}$), then the
tunneling rate grows with the parameter $\lambda$. In other words, in
order to minimize the rate of double-mutant production, one would need
to keep the stem cell pool as small as possible. 

For neutral and advantageous intermediate mutants, where the symmetric
division model gives rise to the lowest double-mutant production rate
compared to the homogeneous model and the hierarchical model with
asymmetric divisions, this rate is minimized for a particular fraction
of stem cells. This fraction is defined by the mutation rate $u_2$ in the neutral
case, and by the fitness of the intermediate mutants in the case of
weakly advantageous mutants.  For neutral one-hit mutants ($|1-r|\ll
\sqrt{u_2}$), the optimal value of $\lambda$ is given by
\begin{equation}\label{eq:opt1}
\lambda_{opt}=1+2u_2^{1/3},
\end{equation}
and for weakly advantageous mutants with $1<r<\lambda$, $|r-1|\gg \sqrt{u_2}$, we have
\begin{equation}\label{eq:opt}
\lambda_{opt}=\frac{r}{2-r}.
\end{equation}
For example, for the biologically most relevant case of neutral
one-hit mutants, the optimal fraction of stem cells is approximately
$1\%$ of the total population, assuming $u_2=10^{-7}$. 

\begin{figure}[!ht]
\begin{center}
\includegraphics[width=6.9in]{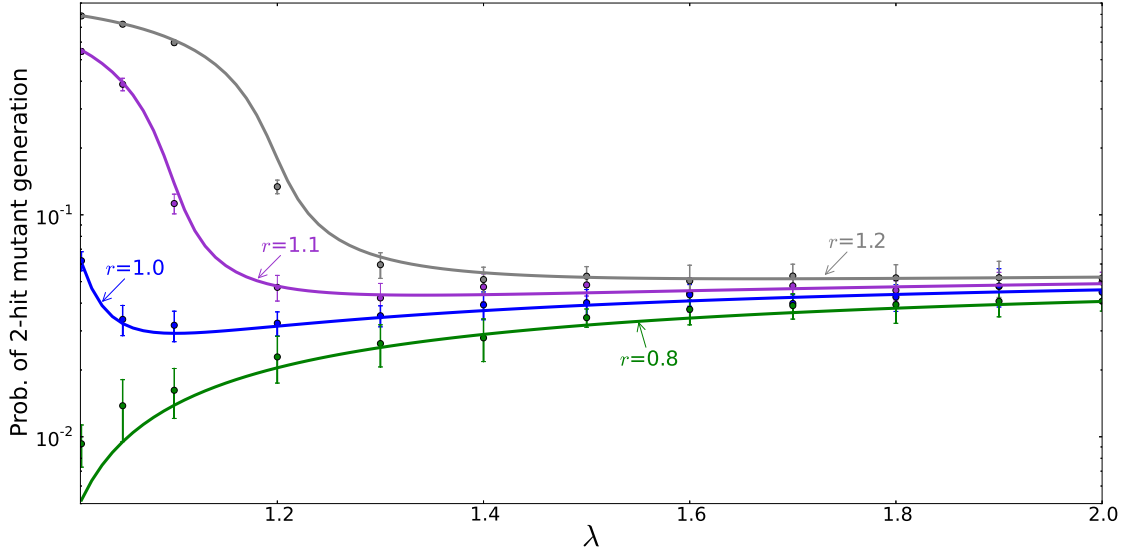}
\end{center}
\caption{
{\bf The probability of double-hit mutant generation in the symmetric division model.} 
The case of symmetrically dividing stem cells, same as in figure \ref{fig:DifferentR}.}
\label{fig:SymDifferentR}
\end{figure}

These results are illustrated in figure \ref{fig:SymDifferentR}.  In
this plot, we can see for $r=0.8$ the probability of having a doubly
mutated cell (after a given time-span) is an increasing function of
$\lambda$, as predicted. For the case of $r=1$, the numerical
simulation in figure \ref{fig:SymDifferentR} shows that
$\lambda_{opt}\approx 1.1$ (compared with $\lambda_{opt}=1.093$
predicted by formula \eqref{eq:opt1}).  For the case $r=1.2$, formula
(\ref{eq:opt}) gives $\lambda_{opt}\approx 1.5$, which approximately
coincides with the numerical optimum. In the case of advantageous
mutants however the minima of $\lambda$ are very shallow.

\subsection*{Do mutations in TA cells produce double-mutants?} 

\begin{figure}[!ht]
\begin{center}
\includegraphics[width=4in]{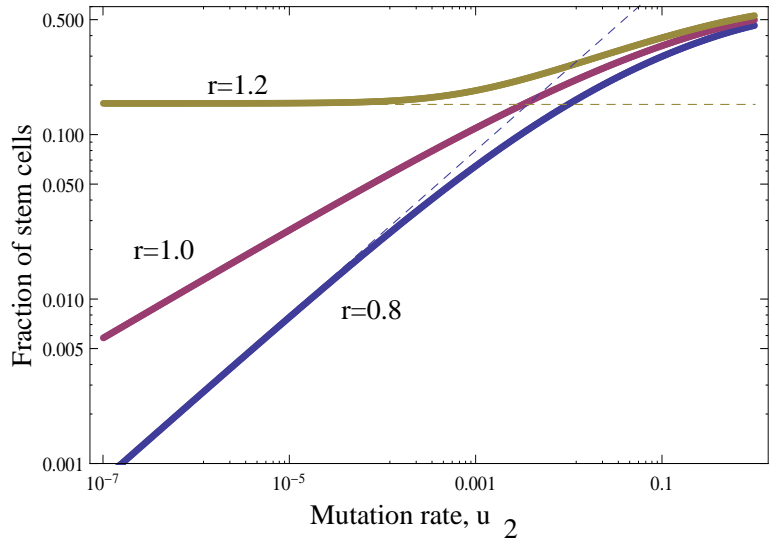}
\end{center}
\caption{
{\bf The threshold fraction of stem cells corresponding to stem and TA cells contributing equally to double-hit mutant production.} The quantity $1-1/\lambda_c$, which corresponds to $R^{stem}_{0\to 2}= R^{TA}_{0\to 2}$, is plotted as a function of the mutation rate, $u_2$,  for three different values of $r$, and $\sigma=1$. For the fraction of stem cells above these values, stem cells have a higher contribution to the rate of double-mutant production compared to the non-stem cells. Thin dashed lines show the approximations of equation (\ref{lam}). }  
\label{fig:lam}
\end{figure}

Let us compare the relative contributions to the double-mutant production rate coming from stem cells and TA cells, equation (\ref{rate}):
\beq
\label{stem_no}
R^{stem}_{0\to 2}=\frac{Nu_1}{2}\left(1-\frac{1}{\lambda}\right)(1-y_*),\quad R^{TA}_{0\to 2}=\frac{Nu_1}{2}\left(1+\frac{1}{\lambda}\right)(1-y).
\eeq
The contribution from the TA cells grows as the fraction of TA cells increases. In figure \ref{fig:lam} we plot the fraction of stem cells (given by $1-1/\lambda$) that corresponds to $R^{stem}_{0\to 2}= R^{TA}_{0\to 2}$. We can see that for the mutation rates around $10^{-7}$, this fraction is about $0.1\%$ for disadvantageous intermediate mutants, about $0.5\%$ for neutral mutants,  and about $15\%$ for advantageous mutants. This means that as long as the fraction of stem cells in the population is lower than these threshold values, TA cells contribute {\it more} to the production of double-hit mutants than stem cells. This threshold fraction grows for  larger mutation rates, making it easier for TA cells to contribute significantly to the double-hit mutant production. An analytical approximation for the threshold value of $\lambda$ can be found for small values of mutation rates, such as 
\beq
\label{lam}
\lambda_c=\left\{\begin{array}{ll}
1+r\sqrt{\frac{2\sigma u_2}{1-r}}, & r<1, \mbox{ Regime (1A),}\\
r-\frac{(r-1)^2}{2\sigma}, & r>1, \mbox{ Regime (1C)}.
\end{array}\right.
\eeq

Next we address the question of optimization assuming that only mutations acquired by stem cells are dangerous and can lead to further malignant transformations. In this case, the rate of mutant production is given by $R^{stem}_{0\to 2}$, equation (\ref{stem_no}). It is easy to show that this quantity is maximized by asymmetric divisions only ($\sigma=0$), and it is minimized by symmetric divisions of stem cells ($\sigma=1$). Thus the message of this paper does not change if only stem cell mutations are assumed to contribute to carcinogenesis.

\section*{Discussion}

In this paper we found that symmetrically dividing stem
cells are characterized by a significantly lower rate of two-hit
mutant generation, compared to asymmetrically-dividing cells. This is
especially important in the context of tumor-suppressor gene
inactivation, which is one of the more common patterns of
carcinogenesis. This provides an evolutionary framework for reasoning about stem cell division patterns. 

In the literature, both types of stem cell divisions have been
reported in various tissues. It has also been reported that the same
stem cells are capable of both symmetric and asymmetric
divisions. Whether a cell divides symmetrically or asymmetrically
depends on factors such as the polarized organization of the dividing
cell as well as the cell cycle length \cite{huttner2005symmetric}. In
Drosophila germ stem cells, cell division is asymmetric or symmetric
depending on whether the orientation of the mitotic spindle is
perpendicular or parallel to the interface between the stem cell and
its niche\cite{yin2006stem}. Similarly, mammalian stem cells have been
reported to employ both symmetric and asymmetric divisions to regulate
their numbers and tissue homeostasis \cite{noctor2004cortical,
  morrison2008stem}. A switch from a symmetric mode of divisions to
the asymmetric model has also been reported to take place in
development (see \cite{egger2011regulating, egger2010notch} in the
context of Drosophila).

The fact that the rate of double-hit mutant production is the lowest
for symmetrically dividing cells does not in itself explain or predict
any aspects of the tissue architecture. It however provides an
alternative hypothesis for the observation that in mammalian tissues,
symmetric patterns of stem cell division seem to be very common. The
force of selection that comes from the cancer-delaying effect of such
an architecture can be thought to have helped shape the observed
division patterns. On the other hand, in more primitive organisms such as
Drosophila, asymmetric stem cell divisions seem to dominate adult
homeostasis (following the predominantly symmetric division patterns
of development). Since cancer delay does not provide an important
selection mechanism in the context of Drosophila, we can argue that
this could help explain the observed differences.

\subsection*{Symmetric divisions can have a cancer-delaying effect}

The mathematical result obtained here is that symmetrically dividing cells appear to delay double-hit mutant production compared to an equivalent system with asymmetrically dividing stem cells. What is the intuition behind this finding? Double-mutants are generated by means of mutations that happen in singly-mutated cells. To understand this process, let us focus on the dynamics of single mutants. In particular, we concentrate on singly-mutated stem cells, because the fates of single mutations in TA cells are identical in the two models. What happens to a singly-mutated stem cell under the different division patterns?

\begin{figure}[!ht]
\begin{center}
\includegraphics[width=5in]{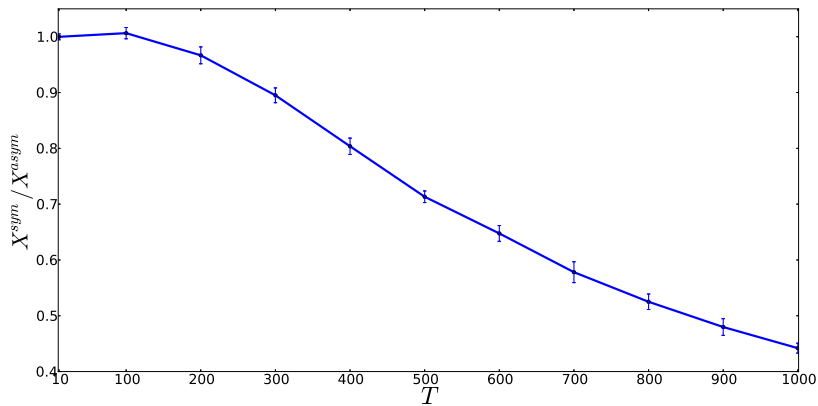}
\end{center}
\caption{
{\bf Why are symmetrically dividing stem cells produce mutants slower?} The weight of a typical symmetrically dividing mutant stem cell lineage, $X^{sym}$, relative to the weight of an asymmetrically dividing mutant stem cell lineage, $X^{asym}=T$, is plotted as a function of the number of stem cell divisions, $T$. Here, $S=20$, $N=1000$, and 20 batches of $10,000$ simulations  were performed to calculate the mean and the standard deviation.} 
\label{fig:exp}
\end{figure}

If stem cells divide asymmetrically, then a mutation acquired in a
stem cell will remain in the system indefinitely, because at every
cell division, a new copy of the mutant stem cell will be
generated. On the other hand, a mutant stem cell generated under the
symmetric division model has a very different and much less certain
fate. Each division of a mutant stem cell can result either in (1)
elimination of the mutation from the stem cell compartment as a result
of a differentiation, or (2) creation of an additional mutant stem
cell as a result of a proliferation event. Superficially, it might
look like the two processes might balance each other out. This
intuition is however misleading. A lineage of mutant stem cells
starting from a single mutant stem cell is much more likely to die out
than to persist and expand. In fact, only $1/K$ of all such lineages
will expand to size $K$. Half of the lineages will differentiate out
after the very first division. Statistically there will be occasional,
rare long-lived lineages, but the vast majority will leave the stem
cell compartment after a small number of divisions. The production of
those "lucky" long-lived mutants is not enough to counter-balance the
great majority of the dead-end lineages that quickly exit the stem
cell compartment. This is illustrated in figure \ref{fig:exp}, which plots the "weight" (the net size of a lineage over time, $T$) of a typical symmetrically dividing mutant stem cell, $X^{sym}$, divided by the weight of a typical asymmetrically dividing mutant stem cell, $X^{asym}$. The latter quantity is simply given by $T$, and the former quantity is a stochastic variable. We can see that the weight of symmetrically dividing mutant lineages is always lower than that of asymmetrically dividing lineages, which means that the former will have a lower probability to produce double-mutants offspring. We conclude that the uncertainty of fate of single mutant stem cells
is the reason for the statistically longer time it takes for
the symmetrically dividing stem cell model to produce a double-hit mutant.

Interestingly, the above argument can be made in a similar manner for
disadvantageous, neutral, or advantageous mutants. In any of those
cases, an asymmetrically dividing mutant stem cell remains in the
population indefinitely. In the model with symmetric divisions,
whenever a mutant stem cell is chosen for division, its probability to
proliferate is similar to its probability to differentiate (in order
to keep the homeostasis), and this the dynamics of each lineage is
independent of its fitness (except that the frequency of updates is
determined by the fitness of mutants; this is why the fitness
parameter $r$ factors out of equations (\ref{Y1}) and (\ref{eqy*1})).

We note that the effect of double-hit mutant production delay caused by symmetric divisions compared to asymmetric divisions is very significant. The difference in the tunneling rate which characterizes the time-scale of the process can be as high as $1,000$-fold for tissues with $10\%$ of stem cells and the mutation rate of $10^{-7}$ per gene per cell division. 

\subsection*{Can TA cells create double-hit mutants?}

The model studied in this paper tracks single- and double-hit mutant
production in both stem and TA cells. It is interesting to
compare which mechanism (through stem cell single mutants or
TA cell single mutants) contributes more to the
double-mutant production? It turns out that as long as the fraction of
stem cells is smaller than a threshold (or equivalently, if the
fraction of the TA cells is larger than a threshold),
non-stem cells contribute equally or more to the production of
double-mutants. This threshold fraction depends on (1) the mutation
rate and (2) the fitness of intermediate, one-hit mutants. For
example, if the intermediate mutants are neutral and the mutation rate
is $10^{-7}$ per gene per cell-division, then the threshold fraction of
stem cells is about $0.5\%$ of the total population. In other words,
mutations originating in non-stem cells are significant if
stem cells comprise less than $0.5\%$ of the total population. This
number is much higher if the intermediate mutants are advantageous, or
if the mutation rate responsible for the second hit is higher. For
$u_2=10^{-3}$, non-stem cells are the driving force behind
double-mutant production as long as stem cells comprise less than
about $10\%$ of the total population. This scenario is realistic in
the presence of genetic instability, where inactivation of a tumor
suppressor gene is likely to occur through a small-scale
mutation of the first copy of the gene followed by a loss of
heterozygocity event inactivating the second copy. The latter can
happen a rate as high as $10^{-2}$ \cite{lengauer1997genetic}. 

The arguments presented above clarify  some aspects of the
long-standing debate about the origins of cancer, see also
\cite{komarova2004initiation}. It is sometime argued that TA cells are
unimportant for cancer initiation, for the following (quantitative)
reason, unrelated to biological evidence. Intuitively, it seems that
 double-hit mutants cannot be created among TA cells, because all
one-hit mutants in the TA compartment will be washed away before they
have a chance to acquire the second hit. As John Cairns writes,
"...there are 256 exponentially multiplying cells that divide twice a
day and are being replenished continually by the divisions of a single
stem cell, none of these 256 cells will ever be separated from the
stem cell by more than eight divisions, and the replication errors
made in those eight divisions are destined, of course, to be
discarded", \cite{cairns2002somatic}. The computations in this paper
demonstrate that under some realistic parameter regimes, double-hit
mutants can be created in the TA compartment, and TA cells
statistically can contribute equally or more to double-hit mutant production
compared to stem cells. The simple reason for this is as follows. Even
though TA cells are short-lived, and getting a second mutation in a
singly-mutated TA cell is unlikely, there are many more TA cells than stem
cells. The low chance of double-mutant generation in a single TA
cells can be outweighed by the fact that TA cells are a large
majority, and single probabilities add up to create a significant
effect.

\subsection*{Cancer stem cell hypothesis}

The question discussed above is purely mathematical, and deals with
the simple possibility to acquire two hits in the TA compartment. A
related biological question is whether mutations occurring in the TA
compartment can lead to further carcinogenic transformations, which
brings us to the cancer stem cell hypothesis \cite{jordan2006cancer,
  nguyen2012cancer}. While the concept of the cancer stem cell remains
controversial \cite{vermeulen2008cancer, gupta2009cancer}, here we do
not intend to argue for or against this theory.  Moreover, we refrain
from making specific interpretations of this theory with regards to
the exact origins of cancer. It has been argued that there is a
distinction between the broader concept of the cancer stem cell on the
one hand, and the narrower concept of normal stem cell becoming
cancerous \cite{nguyen2012cancer}. While the cancer stem cell
hypothesis states that cancer is maintained by a small fraction of
cells with stem-like properties, without making a specific assumption
of how those cells are generated, the more narrow theory argues that
mutations generated among non-stem cells cannot be cancer-initiating,
because (at least, some) cancers originate via the creation of a
cancer stem cell, which is a modified stem cell that retains some
characteristics of "stemness".

In the light of this latter hypothesis, let us analyze the process of double-hit mutant production that occurs via
mutations in stem cells only. Will our results change if only stem cell mutations can lead to carcinogenic transformation? To accommodate this assumption in our model, we must only use the first term in equation (\ref{rate}). It turns out that in this case, the message remains exactly
the same: symmetrically dividing stem cell systems are
characterized by a slower production of double-hit mutants compared to
asymmetrically dividing stem cells. The universality of this result is
explained above: the fate of mutations originating in the
differentiated compartment is identical under the two models, and the
only difference comes from the fates of mutant stem cells.

\subsection*{Stochastic tunneling in the context of hierarchical tissue architecture}

Our theoretical results on the rate of double-hit mutant formation
provide a generalization of a number of previous papers that studied
the process of stochastic tunneling. The concept of stochastic
tunneling was introduced by \cite{nowak2002role, komarova2003mutation}
when studying the first step in colon cancer initiation, the
inactivation of the tumor suppressor gene APC. The concept has later
been investigated by several groups in the context of cancer
initiation, escape dynamics \cite{iwasa2004stochastic}, and more
broadly as a means of crossing an evolutionary valley by an evolving
species \cite{weissman2009rate}. The basic Moran process in a
homogeneous tissue has been used as the underlying mathematical
model. A spatial generalization for the tunneling rate was calculated
in \cite{komarova2006spatial}, and a generalization to a specific
model of renewing epithelial tissue was given in
\cite{komarova2004initiation, komarova2007stochastic}. The present
paper expands the notion of stochastic tunneling to tissues consisting
of stem and differentiated cells, whose fate can vary and is governed
by relatively complex rules. Formula (\ref{rate}) includes the basic
tunneling law of \cite{nowak2002role, komarova2003mutation} as a
special case, and provides a way to predict the rate of mutant
generation based on the stem cell fraction, the mutant fitness, and
the probability of symmetric vs asymmetric divisions.

\bigskip

Finally, we emphasize some of the important simplifications used in
the present model. We considered a two-compartment (stem/TA)
model where all non-stem cells were treated as a single type. Our
numerical explorations suggest that the addition of more compartments does
not change the message of the paper, that is, in the presence of more
cell types, symmetric divisions continue to minimize the rate of
double-mutant production. Further, the effect of the stem cell niche
was modeled in a very basic manner, by assuming the existence of a
stem cell compartment and a relatively tight regulation of
differentiation vs proliferation decisions. Future directions include
the addition of a more detailed description of spatial interactions,
and the inclusion of other cellular processes such as
de-differentiation.

\section*{Methods}

\subsection*{Numerical simulations}

A stochastic numerical simulation was set up according to the
following generalized Moran (constant total population) process. The
population consists of four types of cells: stem cells (wild-type,
$i_*$, and one-hit mutants, $j_*$), and TA cells (wild-type, $i$, and
one-hit mutants, $j$). We have $i+i_*+j+j_*=N$, where $N$ is a
constant total population size. The dynamics proceed as a sequence of
updates. At each update, one TA cell is randomly removed from the
population, and replaced with an offspring of another cell, thus
keeping the total population size constant.

The process of division is modeled as follows. All cells (stem or TA
cells) have a probability to divide. A cell is chosen for division
based on its fitness. The fitness of mutated cells is given by $r$ and
the fitness of wild-type cells is $1$. Let us use the notation ${\cal
  N}=i+i_*+r(j+j_*)$. Then the probability that a wild-type stem cell
is chosen for division is given by $i_*/{\cal N}$; the probability
that a mutated stem cell is chosen for division is given by $rj_*/{\cal
  N}$; the probability that a wild-type TA cell is chosen for division
is given by $i/{\cal N}$; and the probability that a mutant TA cell is
chosen for division is given by $rj/{\cal N}$.

If a wild-type TA cell divides, it creates another wild-type TA cell with
probability $1-u_1$, and it creates a one-hit mutant TA cell with
probability $u_1$. If a mutant TA cell divides, it creates a one-hit
mutant TA cell with probability $1-u_2$, and it creates a two-hit
mutant with probability $u_2$. In case of such an event, the process
stops.

Divisions of stem cells can be either symmetric (with probability
$\sigma$) or asymmetric (with probability $1-\sigma$). Asymmetric
divisions result in a creation of a TA cell. If a wild-type stem cell
is dividing asymmetrically, then with probability $1-u_1$ no mutations
happen, and a one-hit mutant will be created with probability
$u_1$. In case of such an event, with probability $1/2$ the TA
daughter cell will get a mutation, and with probability $1/2$ it will
be the stem cell that acquires a mutation. Similarly, a one-hit mutant
stem cell that divides symmetrically will create a two-hit mutant with
probability $u_2$, in which case the process stops.

Symmetric divisions can be of two types: a differentiation, which
results in a replacement of the dividing stem cell with two TA cells,
or a proliferation which results in a creation of a stem cell. The
probability of proliferation is taken to be
$p=\frac{(i_*+j_*)^{10}}{S^{10}+(i_*+j_*)^{10}}$, where $S$ is a
constant parameter which measures the expected number of stem cells in
the system. The probability of proliferation is given by $1-p$. Again,
when a wild-type stem cell divides, with probability $1-u_1$ both
daughter cells are wild-type, and with probability $u_1$ one of the
daughter cells is a one-hit mutant. If a one-hit mutant stem cell
divides, both daughter cells are one-hit mutants with probability
$1-u_2$, and with probability $u_2$ the process stops because a
double-hit mutant is created.

\begin{figure}[!ht]
\begin{center}
\includegraphics[width=5in]{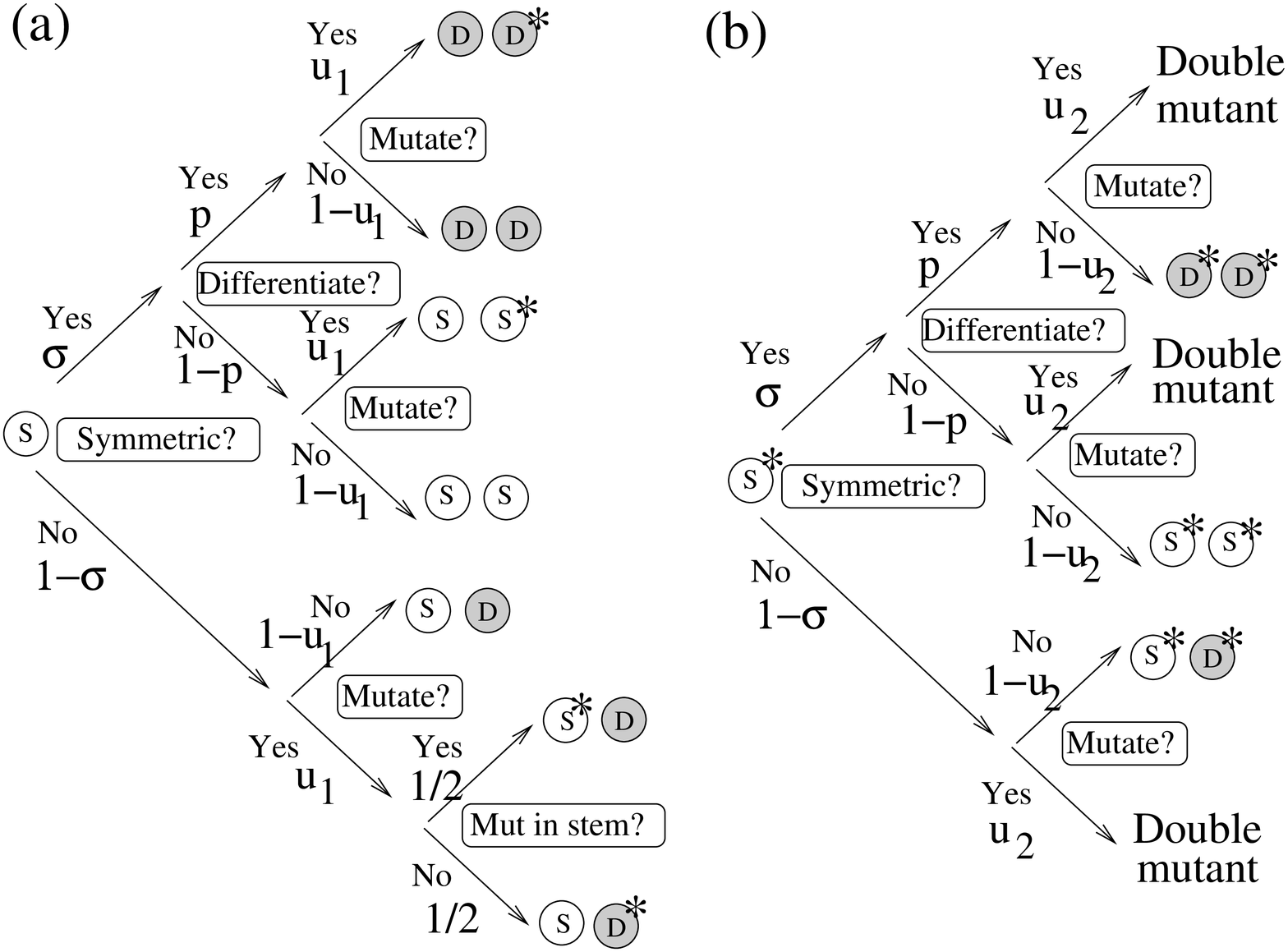}
\end{center}
\caption{
{\bf Stem cell division decision trees for the numerical algorithm.}  (a) Divisions of wild-type stem cells. (b) Divisions of mutant stem cells. Stem cells are denoted by light circles with an ``S'' and TA cells by shaded circles with a ``D''. One-hit mutants are marked with a star. 
}
\label{fig:al}
\end{figure}

The decision trees for stem cells are shown in figure \ref{fig:al}, for wild-type stem cells (a) and for mutated stem cells (b). Stem cells are denoted by light circles with ``S'' and TA cells by shaded circles with ``D''. One-hit mutants are marked with a star. 

These updates were performed repeatedly until either a double-hit
mutant was created, or the maximum number of time-steps was reached,
which was set to $1000$.  We ran this code for 1000 times.  After that
we calculated the fraction of runs that resulted in a double-hit
mutant, which approximates the probability of double-mutant
creation. This quantity was calculated 10 times, and then the averages
and standard deviations were calculated.

To simulate the homogeneous Moran process, the same updates were
performed except the number of stem cells was zero, $i_*+j_*=0$.

\subsection*{Analytical tools}

Suppose we have the following version of the Moran process, which consists of a sequence of elementary updates.  At each update, a daughter
cell is chosen for death at random. Then a cell (a stem cell or a
differentiated cell) is chosen to divide, according to its fitness,
with mutants having fitness $r$. If a differentiated cell is chosen
for division, it divides and this concludes the update. If however a
stem cell is chosen for division, we proceed as follows. (1) With
probability $1-\sigma$, the stem cell can divide asymmetrically, which
concludes this step. (2) With probability $\sigma$, the stem cell
divides symmetrically by differentiation, which is followed by a
proliferation of another randomly chosen stem cell. Finally, another
daughter cell is chosen for death, which concludes this step.

The process described above is slightly different from the numerical
agent-based algorithm outlined used in numerical simulations. In the
generalized Moran process described here, the numbers of stem cells
($S$) and differentiated cells ($D$) are kept constant at every
step. This is a simplification that allowed for analytical
tractability (see below).  In the numerical simulations the number of
stem and differentiated cells fluctuates around a mean value, but
despite this difference, the analytical formulas derived here are in an
excellent agreement with the simulations.

Note that in order to keep $S$ constant, the symmetric stem cell divisions have to come in pairs (one proliferation and one differentiation event), and must be combined with two cell death events. Therefore, on the biological time-scale, an update involving symmetric divisions must have an average duration of two (and not one) elementary updates. Therefore below, when calculating various transition probabilities, the terms associated with symmetric divisions require a factor $1/2$. 

Let us denote by $j_*$ the number of single-mutant stem cells and by $j$ the number of single-mutant differentiated cells. The updates can be envisaged as a Markov process in the space $(j_*,j)$, where $j_*,j\ge 0$, with an additional state $E$ denoting the generation of a double-mutant cell. Below we will use the condition that mutants are drifting at low numbers, $j_*\ll S$ and $j\ll D$. 
We have the following
probabilities:
\begin{itemize}
\item  The probability that the number of mutant differentiated cells increases by one can be approximated as follows: 
$$P_{j_*,j\to j_*,j+1}=\frac{r(j+(1-\sigma)j_*)}{N}(1-u_2)+\frac{D}{N}u_1+(1-\sigma)\frac{S}{N}\frac{u_1}{2}+\frac{u_1S\sigma}{2N},$$
which is (i) the probability that a death of a wild-type differentiated cell (probability $\approx 1$), is followed by either a faithful division of a mutant differentiated cell, or a faithful asymmetric division of a mutant stem cell; (ii)  a division of a wild-type differentiated cell with a mutation; (iii) an asymmetric division of a wild-type stem cell with a mutation happening in the differentiated daughter cell; (iv) a symmetric division of a wild-type stem cell with a mutation (times $1/2$ by association with the symmetric division process). 
\item  The probability that the number of mutant differentiated cells decreases by one: 
$$P_{j_*,j\to j_*,j-1}=\frac{j}{D},$$
which is the probability that a mutant differentiated cell dies followed by a faithful division of any w.t. cell ($\approx 1$). 
\item The probability that the number of mutant differentiated cells increases by two, and the number of mutant stem cells decreases by one:  
$$P_{j_*,j\to j_*-1,j+2}=\frac{\sigma rj_*}{2N}(1-u_2),$$
which is only possible for a symmetric update, when two w.t. differentiated cells die (probability $\approx 1$) followed by a mutant stem cell differentiating without a further mutation (probability $\frac{\sigma rj_*}{N}(1-u_2)$), followed by a w.t. stem cell proliferating without a mutation (probability $\approx 1$); the factor $1/2$ comes from the symmetric update. 
\item  The probability that the number of mutant stem cells increases by one: 
$$P_{j_*,j\to j_*+1,j}=\frac{(1-\sigma)S}{N}\frac{u_1}{2}+\frac{\sigma}{2} \left(\frac{Su_1}{N}+\frac{rj_*}{N}(1-u_2)\right),$$
which is (i) the probability that following a death of a wild-type differentiated cell ($\approx 1$), a wild-type stem cell divides asymmetrically with a mutation in the stem cell daughter cell,  (ii) a wild-type stem cell proliferates with a mutation ($u_1$), or (iii) a mutant stem cell proliferates without a further mutation ($\frac{rj_*}{S}(1-u_2)$). 
\item The probability to create a double-hit mutant: 
$$P_{j_*,j\to E}=\frac{rju_2}{N}+(1-\sigma)\frac{ rj_*u_2}{N}+\sigma \frac{rj_*u_2}{N},$$
which is (i) the probability that a mutant differentiated cell divides with a  mutation, (ii)  a mutant stem cell divides asymmetrically with a mutation, or (iii) a mutant stem cell undergoes either a differentiation or a proliferation event with a mutation. 
\end{itemize}
Let us define by $\varphi_{j_*,j}(t)$ the probability to have $j_*$ mutated stem cells and $j$ mutated differentiated cells at time $t$. The Kolmogorov forward equation for this function is given by
\bbar
\dot \varphi&=&\varphi_{j_*,j-1}\left[r(j-1+(1-\sigma)j_*)(1-u_2)+Du_1+\frac{Su_1}{2}\right]\nonumber \\
&+&\varphi_{j_*,j+1}\lambda (j+1)\nonumber \\
&+&\varphi_{j_*+1,j-2}\frac{\sigma r (j_*+1)}{2}(1-u_2)\nonumber \\
&+&\varphi_{j_*-1,j}\left[Su_1/2+\frac{\sigma}{2}r(j_*-1)(1-u_2)\right]\nonumber \\
&-&\varphi_{j_*,j}(r(j_*+j)+Nu_1+\lambda j).
\eear
Let us define the probability generating function,
$$\Psi(y_*,y;t)=\sum_{j_*,j}\varphi_{j_*,j}(t)y_*^{j_*}y^j.$$
The probability to be in one of the states $(j_*,j)$ is given by $\Psi(1,1;t)$. Therefore, the probability to transit to state $E$ is $P_2(t)=1-\Psi(1,1;t)$. The probability generating function satisfies the following  first order PDE, derived by the standard methods (see e.g. \cite{wodarz2005computational}):
\bbar
\frac{\partial \Psi}{\partial t}&=&
\frac{\partial \Psi}{\partial y_*}\left([\frac{r\sigma}{2}(y_*^2+y^2)+r(1-\sigma)y_*y](1-u_2)-ry_*\right)\nonumber \\
&+&\frac{\partial \Psi}{\partial y}(r(1-u_2)y^2+\lambda(1-y) -ry )\nonumber \\
&-&u_1\Psi \left(\left(D+\frac{S}{2}\right)(1-y)+\frac{S}{2}(1-y_*)\right).
\eear
We have
\beq
\label{P2new}
P_2(t)=1-exp\left(-u_1\int_0^t\left\{\left(D+\frac{S}{2}\right)(1-y(t'))+\frac{S}{2}(1-y_*(t'))\right\}\,dt'\right),
\eeq
where
\bbar
\label{eqy*1}
\dot y_*&=&\left[\frac{r\sigma}{2}(y_*^2+y^2)+r(1-\sigma)y_*y\right](1-u_2)-ry_*,\\
\label{eqy1}
\dot y&=&r(1-u_2)y^2+\lambda(1-y) -ry,\\
y_*(t)&=&y(0)=1.
\eear
Equation (\ref{P2new}) states that one-hit mutants in differentiated cells are produced by divisions of differentiated cells at the rate $u_1D$ and by divisions of stem cells at the rate $u_1S/2$. The factor $1/2$ comes from the fact that in asymmetric divisions, only a half of mutations will be in the differentiated cells, and in symmetric divisions which consist of pairs differentiation/proliferation, only half of the time a mutation will happen upon differentiation. Mutations in stem cells are produced by the divisions of stem cells at rate $u_1S/2$.  The ordinary differential equations describe the dynamics of lineages that start from one differentiated mutant (equation for $\dot y$) or from one stem cell mutant (equation for $\dot y_*$). The dynamics of differentiated mutants is independent of $\sigma$. 

Let us first solve equation (\ref{eqy1}), which informs us about the probability of creating a double-hit mutant in a differentiated cell. This Riccati equation can be solved by standard methods, and the growth of the quantity $1-y$ proceeds in the following stages:
\begin{itemize}
\item The linear growth stage, where $1-y\approx ru_2t$, as long as $t\ll t_*$ (to be defined).
\item The saturation stage, where $1-y\approx C$, as long as $t\gg t_*$. 
\end{itemize}
The constant $C$ obtained from the stable fixed point of equation (\ref{eqy1}) is given by the equation 
$$y=1-\frac{\lambda+r-\sqrt{(\lambda+r)^2-4r\lambda(1-u_2)}}{2r(1-u_2)},$$
and can be approximated by concise expressions as shown below. Given
the solution for $y$, equation (\ref{eqy*1}) can also be analyzed. The
function $1-y_*$ increases monotonically and reaches saturation at
$1-y_*=1$, after characteristic time $t_{**}$. To find that
time-scale, we substitute the constant approximation for the function
$y$, to obtain
$$y_*(t)\approx exp(-t/t_{**}). $$
There are several regimes where the expression take a particularly simple form (see Table \ref{tab:1}). 
\paragraph{Regime (2A).} Let us assume that $|\lambda-r|\gg \sqrt{u_2}$, $r<\lambda$, and $\sigma\ll u_2$. In this case, we have
$$C=\frac{ru_2}{\lambda-r},\quad t_*=\frac{1}{\lambda-r},\quad
t_{**}=\frac{\lambda-r}{\lambda r u_2}\gg t_*.$$
There are therefore three distinct regimes defined by the behavior of the functions $y(t)$ and $y_*(t)$. 
\begin{enumerate}
\item If $t\ll t_*$, we have $1-y=ru_2t$ and $1-y_*=ru_2t+(rt)^2u_2/2$. In this case we have
$$P^{lin}_2(t)=1-exp\left(-\frac{Nu_1u_2rt^2}{2}-\frac{Su_1u_2r^2t^3}{12}\right),$$
where the second term in the exponent is typically smaller than the first, and the behavior is thus indistinguishable for the usual homogeneous Moran process at early times. 
\item If $t_*\ll t\ll t_{**}$, we have $1-y=ru_2/(\lambda-r)$ and $1-y_*=\lambda ru_2t/(\lambda-r)$.  In this case we have
\beq \label{AAt**}
P^{inter}_2(t)=1-exp\left(-\frac{(D+S/2)u_1u_2rt}{\lambda-r}-\frac{Su_1u_2\lambda rt^2}{4(\lambda-r)}\right).
\eeq
\item Finally, if $t\gg t_{**}$, we have $1-y=ru_2/(\lambda-r)$ and $1-y_*=1$, and 
\beq
\label{R02ad}
P^{sat}_2(t)=1-exp\left(-R_{0\to 2}t\right),\quad R_{0\to 2}=\frac{(D+S/2)u_1u_2r}{\lambda-r}+\frac{Su_1}{2}.
\eeq
This regime becomes unimportant if for $t=t_{**}$ we can show that the quantity in the exponent is much larger than one. We have 
$$P^{sat}_2(t_{**})=P^{inter}_2(t_{**})=1-exp\left(-\frac{(D+S/2)u_1}{r}+\frac{Su_1(\lambda-r)}{\lambda ru_2}\right),$$
and this quantity is very close to $1$ for example if $u_1\sim u_2$ and $(\lambda-r)S\gg 1$.
\end{enumerate}

\paragraph{Regime (2B).} Let us assume that $|\lambda-r|\ll \sqrt{u_2}$ and $\sigma\ll \sqrt{u_2}$. In this case, we have
$$C=\sqrt{u_2},\quad t_*=\frac{1}{2\lambda\sqrt{u_2}},\quad t_{**}=\frac{1}{\lambda \sqrt{u_2}}\sim t_*.$$
There are therefore only two regimes defined by the behavior of the functions $y(t)$ and $y_*(t)$. 
\begin{enumerate}
\item If $t\ll t_*$, we have as in the previous case, $1-y=ru_2t$ and $1-y_*=ru_2t+(rt)^2u_2/2$. The probability of double-hit mutant production is thus given by 
$$P^{lin}_2(t)=1-exp\left(-\frac{Nu_1u_2\lambda t^2}{2}-\frac{Su_1u_2\lambda ^2t^3}{12}\right),$$
where the second term in the exponent is typically smaller than the first, and the behavior is thus indistinguishable for the usual homogeneous Moran process at early times. 
\item If $t_*\ll t$, we have $1-y=\sqrt{u_2}$ and $1-y_*=1$.  In this case we have
\beq
\label{R02an}
P^{sat}_2(t)=1-exp\left(-R_{0\to 2}t\right),\quad R_{0\to 2}=(D+S/2)u_1\sqrt{u_2}+\frac{Su_1}{2}.
\eeq
\end{enumerate}

\paragraph{Regime (1A).} Let us 
assume that $|\lambda-r|\gg \sqrt{u_2}$,
$r<\lambda$, and  $\sigma\gg u_2$. The quantity $1-y_*(t)$ behaves as a linear function,
$$1-y_*(t)=\frac{\lambda ru_2t}{\lambda -r},$$
for $t_*\ll t\le t_{**}$, where 
\beq
\label{t**}
t_*=\frac{1}{\lambda-r},\quad t_{**}=\frac{1}{r}\sqrt{\frac{\lambda-r}{2u_2\lambda\sigma}}.
\eeq
For $t\gg t_{**}$, the quantity $1-y_*(t)$ tends to a constant,
\beq
\label{d}
1-y_*(t)=\sqrt{\frac{2u_2\lambda}{\sigma(\lambda-r)}}.
\eeq
Note that the initial behavior of the function $1-y_*(t)$ does not
depend on $\sigma$. This means that for relatively short times ($t\ll
t_{**}$), the mutant generation in stem cells proceeds in the same way
for symmetric and asymmetric divisions. The length of this regime and
the level of saturation however are both functions of $\sigma$. It is
easy to see that both $t_{**}$ and the saturation level increase as
$\sigma$ decreases. This means that the rate of mutant accumulation
becomes higher for asymmetric divisions.

\paragraph{Regime (1B). } Let us assume that $|\lambda-r|\ll \sqrt{u_2}$ and $\sigma\gg \sqrt{u_2}$. Now, the linear stage for $1-y_*(t)$ is defined as
$$1-y_*(t)=r\sqrt{u_2}t,$$
and it occurs for the times $t_*\ll t\ll t_{**}$, where 
$$t_*=\frac{1}{2\lambda\sqrt{u_2}},\quad t_{**}=\frac{1}{\lambda\sqrt{2\sigma}u_2^{1/4}}.$$
For $t\gg t_{**}$, the quantity $1-y_*(t)$ tends to a constant,
\beq
\label{d1}
1-y_*(t)=\sqrt{\frac{2}{\sigma}}u_2^{1/4}.
\eeq

Calculations for regimes (1C) and (2C) are performed in a similar manner, see Table \ref{tab:2}. 

% Do NOT remove this, even if you are not including acknowledgments
\section*{Acknowledgments}

%\section*{Author contributions}

%NK designed study, NK and LS performed the research and wrote the paper. 

%\section*{Conflicts of interest}

%The authors declare no conflict of interest. 

%The support of NIH grant 1R01CA129286-01A1 is gratefully acknowledged. 

%\section*{References}
% The bibtex filename
%\bibliography{template}
\bibliography{symm2}

\begin{thebibliography}{10}
\providecommand{\url}[1]{\texttt{#1}}
\providecommand{\urlprefix}{URL }
\expandafter\ifx\csname urlstyle\endcsname\relax
  \providecommand{\doi}[1]{doi:\discretionary{}{}{}#1}\else
  \providecommand{\doi}{doi:\discretionary{}{}{}\begingroup
  \urlstyle{rm}\Url}\fi
\providecommand{\bibAnnoteFile}[1]{%
  \IfFileExists{#1}{\begin{quotation}\noindent\textsc{Key:} #1\\
  \textsc{Annotation:}\ \input{#1}\end{quotation}}{}}
\providecommand{\bibAnnote}[2]{%
  \begin{quotation}\noindent\textsc{Key:} #1\\
  \textsc{Annotation:}\ #2\end{quotation}}
\providecommand{\eprint}[2][]{\url{#2}}

\bibitem{morrison2006asymmetric}
Morrison SJ, Kimble J (2006) Asymmetric and symmetric stem-cell divisions in
  development and cancer.
\newblock Nature 441: 1068--1074.
\bibAnnoteFile{morrison2006asymmetric}

\bibitem{shen2004endothelial}
Shen Q, Goderie SK, Jin L, Karanth N, Sun Y, et~al. (2004) Endothelial cells
  stimulate self-renewal and expand neurogenesis of neural stem cells.
\newblock Science 304: 1338--1340.
\bibAnnoteFile{shen2004endothelial}

\bibitem{knoblich2008mechanisms}
Knoblich JA (2008) Mechanisms of asymmetric stem cell division.
\newblock Cell 132: 583--597.
\bibAnnoteFile{knoblich2008mechanisms}

\bibitem{fuchs2004socializing}
Fuchs E, Tumbar T, Guasch G (2004) Socializing with the neighbors: stem cells
  and their niche.
\newblock Cell 116: 769--778.
\bibAnnoteFile{fuchs2004socializing}

\bibitem{zhong2008neurogenesis}
Zhong W, Chia W (2008) Neurogenesis and asymmetric cell division.
\newblock Current opinion in neurobiology 18: 4--11.
\bibAnnoteFile{zhong2008neurogenesis}

\bibitem{ho2005kinetics}
Ho AD (2005) Kinetics and symmetry of divisions of hematopoietic stem cells.
\newblock Experimental hematology 33: 1--8.
\bibAnnoteFile{ho2005kinetics}

\bibitem{zhang2009distinct}
Zhang YV, Cheong J, Ciapurin N, McDermitt DJ, Tumbar T (2009) Distinct
  self-renewal and differentiation phases in the niche of infrequently dividing
  hair follicle stem cells.
\newblock Cell Stem Cell 5: 267--278.
\bibAnnoteFile{zhang2009distinct}

\bibitem{loeffler2002tissue}
Loeffler M, Roeder I (2002) Tissue stem cells: definition, plasticity,
  heterogeneity, self-organization and models--a conceptual approach.
\newblock Cells Tissues Organs 171: 8--26.
\bibAnnoteFile{loeffler2002tissue}

\bibitem{marshman2002intestinal}
Marshman E, Booth C, Potten CS (2002) The intestinal epithelial stem cell.
\newblock Bioessays 24: 91--98.
\bibAnnoteFile{marshman2002intestinal}

\bibitem{clayton2007single}
Clayton E, Doup{\'e} DP, Klein AM, Winton DJ, Simons BD, et~al. (2007) A single
  type of progenitor cell maintains normal epidermis.
\newblock Nature 446: 185--189.
\bibAnnoteFile{clayton2007single}

\bibitem{liu2000loss}
Liu M, Pleasure S, Collins A, Noebels J, Naya F, et~al. (2000) Loss of
  beta2/neurod leads to malformation of the dentate gyrus and epilepsy.
\newblock Proceedings of the National Academy of Sciences 97: 865--870.
\bibAnnoteFile{liu2000loss}

\bibitem{simmons2003cyclic}
Simmons C, Matlis S, Thornton A, Chen S, Wang C, et~al. (2003) Cyclic strain
  enhances matrix mineralization by adult human mesenchymal stem cells via the
  extracellular signal-regulated kinase (erk1/2) signaling pathway.
\newblock Journal of biomechanics 36: 1087--1096.
\bibAnnoteFile{simmons2003cyclic}

\bibitem{alvarez2004long}
Alvarez-Buylla A, Lim D (2004) For the long run: maintaining germinal niches in
  the adult brain.
\newblock Neuron 41: 683--686.
\bibAnnoteFile{alvarez2004long}

\bibitem{saha2006inhibition}
Saha S, Ji L, de~Pablo J, Palecek S (2006) Inhibition of human embryonic stem
  cell differentiation by mechanical strain.
\newblock Journal of cellular physiology 206: 126--137.
\bibAnnoteFile{saha2006inhibition}

\bibitem{lien2006alpha}
Lien W, Klezovitch O, Fernandez T, Delrow J, Vasioukhin V (2006) $\{$alpha$\}$
  e-catenin controls cerebral cortical size by regulating the hedgehog
  signaling pathway.
\newblock Science's STKE 311: 1609.
\bibAnnoteFile{lien2006alpha}

\bibitem{adams2007niche}
Adams G, Scadden D (2007) A niche opportunity for stem cell therapeutics.
\newblock Gene therapy 15: 96--99.
\bibAnnoteFile{adams2007niche}

\bibitem{dehay2007cell}
Dehay C, Kennedy H (2007) Cell-cycle control and cortical development.
\newblock Nature Reviews Neuroscience 8: 438--450.
\bibAnnoteFile{dehay2007cell}

\bibitem{orford2008deconstructing}
Orford K, Scadden D (2008) Deconstructing stem cell self-renewal: genetic
  insights into cell-cycle regulation.
\newblock Nature Reviews Genetics 9: 115--128.
\bibAnnoteFile{orford2008deconstructing}

\bibitem{nusse2008wnt}
Nusse R (2008) Wnt signaling and stem cell control.
\newblock Cell research 18: 523--527.
\bibAnnoteFile{nusse2008wnt}

\bibitem{spiegel2008stem}
Spiegel A, Kalinkovich A, Shivtiel S, Kollet O, Lapidot T (2008) Stem cell
  regulation via dynamic interactions of the nervous and immune systems with
  the microenvironment.
\newblock Cell Stem Cell 3: 484--492.
\bibAnnoteFile{spiegel2008stem}

\bibitem{saha2008tgf}
Saha S, Ji L, De~Pablo J, Palecek S (2008) Tgf [beta]/activin/nodal pathway in
  inhibition of human embryonic stem cell differentiation by mechanical strain.
\newblock Biophysical journal 94: 4123--4133.
\bibAnnoteFile{saha2008tgf}

\bibitem{sen2008mechanical}
Sen B, Xie Z, Case N, Ma M, Rubin C, et~al. (2008) Mechanical strain inhibits
  adipogenesis in mesenchymal stem cells by stimulating a durable
  $\beta$-catenin signal.
\newblock Endocrinology 149: 6065--6075.
\bibAnnoteFile{sen2008mechanical}

\bibitem{guilak2009control}
Guilak F, Cohen D, Estes B, Gimble J, Liedtke W, et~al. (2009) Control of stem
  cell fate by physical interactions with the extracellular matrix.
\newblock Cell Stem Cell 5: 17--26.
\bibAnnoteFile{guilak2009control}

\bibitem{lavado2010prox1}
Lavado A, Lagutin O, Chow L, Baker S, Oliver G (2010) Prox1 is required for
  granule cell maturation and intermediate progenitor maintenance during brain
  neurogenesis.
\newblock PLoS biology 8: e1000460.
\bibAnnoteFile{lavado2010prox1}

\bibitem{de2010regulation}
de~Graaf C, Kauppi M, Baldwin T, Hyland C, Metcalf D, et~al. (2010) Regulation
  of hematopoietic stem cells by their mature progeny.
\newblock Proceedings of the National Academy of Sciences 107: 21689--21694.
\bibAnnoteFile{de2010regulation}

\bibitem{li2010coexistence}
Li L, Clevers H (2010) Coexistence of quiescent and active adult stem cells in
  mammals.
\newblock Science 327: 542--545.
\bibAnnoteFile{li2010coexistence}

\bibitem{salomoni2010cell}
Salomoni P, Calegari F (2010) Cell cycle control of mammalian neural stem
  cells: putting a speed limit on g1.
\newblock Trends in cell biology 20: 233--243.
\bibAnnoteFile{salomoni2010cell}

\bibitem{hsieh2012orchestrating}
Hsieh J (2012) Orchestrating transcriptional control of adult neurogenesis.
\newblock Genes \& Development 26: 1010--1021.
\bibAnnoteFile{hsieh2012orchestrating}

\bibitem{ordonez2012lrig1}
Ord{\'o}{\~n}ez-Mor{\'a}n P, Huelsken J (2012) Lrig1: a new master regulator of
  epithelial stem cells.
\newblock The EMBO Journal .
\bibAnnoteFile{ordonez2012lrig1}

\bibitem{yatabe2001investigating}
Yatabe Y, Tavar{\'e} S, Shibata D (2001) Investigating stem cells in human
  colon by using methylation patterns.
\newblock Proceedings of the National Academy of Sciences 98: 10839--10844.
\bibAnnoteFile{yatabe2001investigating}

\bibitem{spradling2001stem}
Spradling A, Drummond-Barbosa D, Kai T, et~al. (2001) Stem cells find their
  niche.
\newblock NATURE-LONDON- : 98--104.
\bibAnnoteFile{spradling2001stem}

\bibitem{nicolas2007stem}
Nicolas P, Kim KM, Shibata D, Tavar{\'e} S (2007) The stem cell population of
  the human colon crypt: analysis via methylation patterns.
\newblock PLoS computational biology 3: e28.
\bibAnnoteFile{nicolas2007stem}

\bibitem{campbell1996post}
Campbell F, Williams G, Appleton M, Dixon M, Harris M, et~al. (1996)
  Post-irradiation somatic mutation and clonal stabilisation time in the human
  colon.
\newblock Gut 39: 569--573.
\bibAnnoteFile{campbell1996post}

\bibitem{klein2011universal}
Klein AM, Simons BD (2011) Universal patterns of stem cell fate in cycling
  adult tissues.
\newblock Development 138: 3103--3111.
\bibAnnoteFile{klein2011universal}

\bibitem{klein2010mouse}
Klein AM, Nakagawa T, Ichikawa R, Yoshida S, Simons BD (2010) Mouse germ line
  stem cells undergo rapid and stochastic turnover.
\newblock Cell Stem Cell 7: 214--224.
\bibAnnoteFile{klein2010mouse}

\bibitem{lopez2010intestinal}
Lopez-Garcia C, Klein AM, Simons BD, Winton DJ (2010) Intestinal stem cell
  replacement follows a pattern of neutral drift.
\newblock Science 330: 822--825.
\bibAnnoteFile{lopez2010intestinal}

\bibitem{snippert2010intestinal}
Snippert HJ, van~der Flier LG, Sato T, van Es JH, van~den Born M, et~al. (2010)
  Intestinal crypt homeostasis results from neutral competition between
  symmetrically dividing lgr5 stem cells.
\newblock Cell 143: 134--144.
\bibAnnoteFile{snippert2010intestinal}

\bibitem{simons2011strategies}
Simons BD, Clevers H (2011) Strategies for homeostatic stem cell self-renewal
  in adult tissues.
\newblock Cell 145: 851--862.
\bibAnnoteFile{simons2011strategies}

\bibitem{doupe2010ordered}
Doup{\'e} DP, Klein AM, Simons BD, Jones PH (2010) The ordered architecture of
  murine ear epidermis is maintained by progenitor cells with random fate.
\newblock Developmental cell 18: 317--323.
\bibAnnoteFile{doupe2010ordered}

\bibitem{reya2005wnt}
Reya T, Clevers H (2005) Wnt signalling in stem cells and cancer.
\newblock Nature 434: 843--850.
\bibAnnoteFile{reya2005wnt}

\bibitem{clarke2003regulation}
Clarke RB, Anderson E, Howell A, Potten CS (2003) Regulation of human breast
  epithelial stem cells.
\newblock Cell proliferation 36: 45--58.
\bibAnnoteFile{clarke2003regulation}

\bibitem{caussinus2005induction}
Caussinus E, Gonzalez C (2005) Induction of tumor growth by altered stem-cell
  asymmetric division in drosophila melanogaster.
\newblock Nature genetics 37: 1125--1129.
\bibAnnoteFile{caussinus2005induction}

\bibitem{caussinus2007asymmetric}
Caussinus E, Hirth F (2007) Asymmetric stem cell division in development and
  cancer.
\newblock Asymmetric Cell Division : 205--225.
\bibAnnoteFile{caussinus2007asymmetric}

\bibitem{Aparicio2009}
Aparicio S, Eaves C (2009) p53: a new kingpin in the stem cell arena.
\newblock Cell 138: 1060--2.
\bibAnnoteFile{Aparicio2009}

\bibitem{gonzalez2013drosophila}
Gonzalez C (2013) Drosophila melanogaster: a model and a tool to investigate
  malignancy and identify new therapeutics.
\newblock Nature Reviews Cancer 13: 172--183.
\bibAnnoteFile{gonzalez2013drosophila}

\bibitem{vogelstein2002genetic}
Vogelstein B, Kinzler KW (2002) The genetic basis of human cancer, volume 821.
\newblock McGraw-Hill, Medical Pub. Division New York.
\bibAnnoteFile{vogelstein2002genetic}

\bibitem{knudson1971mutation}
Knudson AG (1971) Mutation and cancer: statistical study of retinoblastoma.
\newblock Proceedings of the National Academy of Sciences 68: 820--823.
\bibAnnoteFile{knudson1971mutation}

\bibitem{knudson2001two}
Knudson AG (2001) Two genetic hits (more or less) to cancer.
\newblock Nature Reviews Cancer 1: 157--162.
\bibAnnoteFile{knudson2001two}

\bibitem{pmid7479951}
Tomlinson IP, Bodmer WF (1995) {{F}ailure of programmed cell death and
  differentiation as causes of tumors: some simple mathematical models}.
\newblock Proc Natl Acad Sci USA 92: 11130--11134.
\bibAnnoteFile{pmid7479951}

\bibitem{pmid17049944}
d'Onofrio A, Tomlinson IP (2007) {{A} nonlinear mathematical model of cell
  turnover, differentiation and tumorigenesis in the intestinal crypt}.
\newblock J Theor Biol 244: 367--374.
\bibAnnoteFile{pmid17049944}

\bibitem{pmid17360468}
Johnston MD, Edwards CM, Bodmer WF, Maini PK, Chapman SJ (2007) {{M}athematical
  modeling of cell population dynamics in the colonic crypt and in colorectal
  cancer}.
\newblock Proc Natl Acad Sci USA 104: 4008--4013.
\bibAnnoteFile{pmid17360468}

\bibitem{pmid18451157}
Boman BM, Fields JZ, Cavanaugh KL, Guetter A, Runquist OA (2008) {{H}ow
  dysregulated colonic crypt dynamics cause stem cell overpopulation and
  initiate colon cancer}.
\newblock Cancer Res 68: 3304--3313.
\bibAnnoteFile{pmid18451157}

\bibitem{pmid12101397}
Hardy K, Stark J (2002) {{M}athematical models of the balance between apoptosis
  and proliferation}.
\newblock Apoptosis 7: 373--381.
\bibAnnoteFile{pmid12101397}

\bibitem{pmid11517339}
Yatabe Y, Tavare S, Shibata D (2001) {{I}nvestigating stem cells in human colon
  by using methylation patterns}.
\newblock Proc Natl Acad Sci USA 98: 10839--10844.
\bibAnnoteFile{pmid11517339}

\bibitem{ganguly2006mathematical}
Ganguly R, Puri I (2006) Mathematical model for the cancer stem cell
  hypothesis.
\newblock Cell proliferation 39: 3--14.
\bibAnnoteFile{ganguly2006mathematical}

\bibitem{ganguly2007mathematical}
Ganguly R, Puri I (2007) Mathematical model for chemotherapeutic drug efficacy
  in arresting tumour growth based on the cancer stem cell hypothesis.
\newblock Cell proliferation 40: 338--354.
\bibAnnoteFile{ganguly2007mathematical}

\bibitem{boman2007symmetric}
Boman BM, Wicha MS, Fields JZ, Runquist OA (2007) Symmetric division of cancer
  stem cells--a key mechanism in tumor growth that should be targeted in future
  therapeutic approaches.
\newblock Clinical Pharmacology \& Therapeutics 81: 893--898.
\bibAnnoteFile{boman2007symmetric}

\bibitem{ashkenazi2008pathways}
Ashkenazi R, Gentry SN, Jackson TL (2008) Pathways to tumorigenesis—modeling
  mutation acquisition in stem cells and their progeny.
\newblock Neoplasia (New York, NY) 10: 1170.
\bibAnnoteFile{ashkenazi2008pathways}

\bibitem{michor2008mathematical}
Michor F (2008) Mathematical models of cancer stem cells.
\newblock Journal of Clinical Oncology 26: 2854--2861.
\bibAnnoteFile{michor2008mathematical}

\bibitem{tomasetti2010role}
Tomasetti C, Levy D (2010) Role of symmetric and asymmetric division of stem
  cells in developing drug resistance.
\newblock Proceedings of the National Academy of Sciences 107: 16766--16771.
\bibAnnoteFile{tomasetti2010role}

\bibitem{enderling2011cancer}
Enderling H, Hahnfeldt P (2011) Cancer stem cells in solid tumors: Is 'evading
  apoptosis' a hallmark of cancer?
\newblock Progress in Biophysics and Molecular Biology .
\bibAnnoteFile{enderling2011cancer}

\bibitem{enderling2007mathematical}
Enderling H, Chaplain MA, Anderson AR, Vaidya JS (2007) A mathematical model of
  breast cancer development, local treatment and recurrence.
\newblock Journal of theoretical biology 246: 245--259.
\bibAnnoteFile{enderling2007mathematical}

\bibitem{enderling2009paradoxical}
Enderling H, Anderson AR, Chaplain MA, Beheshti A, Hlatky L, et~al. (2009)
  Paradoxical dependencies of tumor dormancy and progression on basic cell
  kinetics.
\newblock Cancer research 69: 8814--8821.
\bibAnnoteFile{enderling2009paradoxical}

\bibitem{enderling2009importance}
Enderling H, Park D, Hlatky L, Hahnfeldt P (2009) The importance of spatial
  distribution of stemness and proliferation state in determining tumor
  radioresponse.
\newblock Math Model Nat Phenom 4: 117--133.
\bibAnnoteFile{enderling2009importance}

\bibitem{enderling2009migration}
Enderling H, Hlatky L, Hahnfeldt P (2009) Migration rules: tumours are
  conglomerates of self-metastases.
\newblock British journal of cancer 100: 1917--1925.
\bibAnnoteFile{enderling2009migration}

\bibitem{piotrowska2008mathematical}
Piotrowska M, Enderling H, van~der Heiden U, Mackey M (2008) Mathematical
  modeling of stem cells related to cancer.
\newblock Complex Systems in Biomedicine .
\bibAnnoteFile{piotrowska2008mathematical}

\bibitem{frank2003cell}
Frank SA, Nowak MA (2003) Cell biology: Developmental predisposition to cancer.
\newblock Nature 422: 494--494.
\bibAnnoteFile{frank2003cell}

\bibitem{michor2003stochastic}
Michor F, Nowak MA, Frank SA, Iwasa Y (2003) Stochastic elimination of cancer
  cells.
\newblock Proceedings of the Royal Society of London Series B: Biological
  Sciences 270: 2017--2024.
\bibAnnoteFile{michor2003stochastic}

\bibitem{dingli2007symmetric}
Dingli D, Traulsen A, Michor F (2007) (a) symmetric stem cell replication and
  cancer.
\newblock PLoS computational biology 3: e53.
\bibAnnoteFile{dingli2007symmetric}

\bibitem{nowak2002role}
Nowak MA, Komarova NL, Sengupta A, Jallepalli PV, Shih IM, et~al. (2002) The
  role of chromosomal instability in tumor initiation.
\newblock Proceedings of the National Academy of Sciences 99: 16226--16231.
\bibAnnoteFile{nowak2002role}

\bibitem{komarova2003mutation}
Komarova NL, Sengupta A, Nowak MA (2003) Mutation--selection networks of cancer
  initiation: tumor suppressor genes and chromosomal instability.
\newblock Journal of theoretical biology 223: 433--450.
\bibAnnoteFile{komarova2003mutation}

\bibitem{iwasa2004stochastic}
Iwasa Y, Michor F, Nowak MA (2004) Stochastic tunnels in evolutionary dynamics.
\newblock Genetics 166: 1571--1579.
\bibAnnoteFile{iwasa2004stochastic}

\bibitem{weissman2009rate}
Weissman DB, Desai MM, Fisher DS, Feldman MW (2009) The rate at which asexual
  populations cross fitness valleys.
\newblock Theoretical population biology 75: 286--300.
\bibAnnoteFile{weissman2009rate}

\bibitem{huttner2005symmetric}
Huttner WB, Kosodo Y (2005) Symmetric versus asymmetric cell division during
  neurogenesis in the developing vertebrate central nervous system.
\newblock Current opinion in cell biology 17: 648--657.
\bibAnnoteFile{huttner2005symmetric}

\bibitem{yin2006stem}
Yin T, Li L, et~al. (2006) The stem cell niches in bone.
\newblock Journal of Clinical Investigation 116: 1195.
\bibAnnoteFile{yin2006stem}

\bibitem{noctor2004cortical}
Noctor SC, Mart{\'\i}nez-Cerde{\~n}o V, Ivic L, Kriegstein AR (2004) Cortical
  neurons arise in symmetric and asymmetric division zones and migrate through
  specific phases.
\newblock Nature neuroscience 7: 136--144.
\bibAnnoteFile{noctor2004cortical}

\bibitem{morrison2008stem}
Morrison SJ, Spradling AC (2008) Stem cells and niches: mechanisms that promote
  stem cell maintenance throughout life.
\newblock Cell 132: 598--611.
\bibAnnoteFile{morrison2008stem}

\bibitem{egger2011regulating}
Egger B, Gold KS, Brand AH (2011) Regulating the balance between symmetric and
  asymmetric stem cell division in the developing brain.
\newblock Fly 5: 237--241.
\bibAnnoteFile{egger2011regulating}

\bibitem{egger2010notch}
Egger B, Gold KS, Brand AH (2010) Notch regulates the switch from symmetric to
  asymmetric neural stem cell division in the drosophila optic lobe.
\newblock Development 137: 2981--2987.
\bibAnnoteFile{egger2010notch}

\bibitem{lengauer1997genetic}
Lengauer C, Kinzler K, Vogelstein B (1997) Genetic instability in colorectal
  cancers.
\newblock Nature 386: 623--627.
\bibAnnoteFile{lengauer1997genetic}

\bibitem{komarova2004initiation}
Komarova NL, Wang L (2004) Initiation of colorectal cancer: where do the two
  hits hit?
\newblock Cell Cycle 3: 1558--1565.
\bibAnnoteFile{komarova2004initiation}

\bibitem{cairns2002somatic}
Cairns J (2002) Somatic stem cells and the kinetics of mutagenesis and
  carcinogenesis.
\newblock Proceedings of the National Academy of Sciences 99: 10567--10570.
\bibAnnoteFile{cairns2002somatic}

\bibitem{jordan2006cancer}
Jordan CT, Guzman ML, Noble M (2006) Cancer stem cells.
\newblock New England Journal of Medicine 355: 1253--1261.
\bibAnnoteFile{jordan2006cancer}

\bibitem{nguyen2012cancer}
Nguyen LV, Vanner R, Dirks P, Eaves CJ (2012) Cancer stem cells: an evolving
  concept.
\newblock Nature Reviews Cancer 12: 133--143.
\bibAnnoteFile{nguyen2012cancer}

\bibitem{vermeulen2008cancer}
Vermeulen L, Sprick M, Kemper K, Stassi G, Medema J (2008) Cancer stem
  cells--old concepts, new insights.
\newblock Cell Death \& Differentiation 15: 947--958.
\bibAnnoteFile{vermeulen2008cancer}

\bibitem{gupta2009cancer}
Gupta PB, Chaffer CL, Weinberg RA (2009) Cancer stem cells: mirage or reality?
\newblock Nature medicine 15: 1010--1012.
\bibAnnoteFile{gupta2009cancer}

\bibitem{komarova2006spatial}
Komarova NL (2006) Spatial stochastic models for cancer initiation and
  progression.
\newblock Bulletin of mathematical biology 68: 1573--1599.
\bibAnnoteFile{komarova2006spatial}

\bibitem{komarova2007stochastic}
Komarova NL (2007) Stochastic modeling of loss-and gain-of-function mutations
  in cancer.
\newblock Mathematical Models and Methods in Applied Sciences 17: 1647--1673.
\bibAnnoteFile{komarova2007stochastic}

\bibitem{wodarz2005computational}
Wodarz D, Komarova NL (2005) Computational biology of cancer: lecture notes and
  mathematical modeling.
\newblock World Scientific Publishing Company.
\bibAnnoteFile{wodarz2005computational}

\end{thebibliography}

\newpage

%\section*{Figure Legends}

%\newpage

%\section*{Tables}

\end{document}